\documentclass[twocolumn,secnumarabic,amssymb, nobibnotes, aps, prd]{revtex4}
\usepackage{graphicx}

\begin{document}
\title{Observation of the Little-Parks Oscillations in a System of Asymmetric Superconducting Rings}
\author{A. A. Burlakov, V. L. Gurtovoi, S. V. Dubonos, A. V. Nikulov, and V. A. Tulin}
\affiliation{Institute of Microelectronics Technology and High Purity Materials, Russian Academy of Sciences, 142432 Chernogolovka, Moscow District, RUSSIA.} 
\begin{abstract} Little-Parks oscillations are observed in a system of 110 series-connected aluminum rings $2 \ \mu m$ in diameter with the use of measuring currents from $10 \ nA$ to $1  \ \mu A$. The measurements show that the amplitude and character of the oscillations are independent of the relation between the measuring current and the amplitude of the persistent current. By using asymmetric rings, it is demonstrated that the persistent current has clockwise or contra-clockwise direction. This means that the total current in one of the semi-rings may be directed against the electric field at measurement of  the Little-Parks oscillations. The measurements at zero and low measuring current have revealed that the persistent current, like the conventional circulating current, causes a potential difference on the semi-rings with different cross sections in spite of the absence of the Faraday's voltage.
 \end{abstract}

\maketitle

\narrowtext

The Little-Parks experiment \cite{LP1962}, equally with the works \cite{FlQu1961}, is among the first observations of the quantization effects in superconductors, which follow from the Ginzburg-Landau theory \cite{GL1950} and the fluxoid quantization postulated by London \cite{London50} for explaining the Meissner effect. According to the second equation of the Ginzburg-Landau theory \cite{Tinkham}
$$j_{c} = \frac{2e}{m}|\Psi |^{2}(\hbar \nabla \varphi -2eA) = 2e|\Psi |^{2}v \eqno{(1)}$$
and the requirement that the complex pair wave function $\Psi  = |\Psi |exp(i\varphi )$ must be single-valued $\oint dl\nabla \varphi  = 2\pi n$  the superconducting current density $j_{s}$ and the pair velocity $v$ along a closed contour $l$, 
$$\oint dl v = \frac{2e}{m}(n\Phi _{0} - \Phi ) \eqno{(2)}$$
cannot be equal to zero if the pair density $|\Psi |^{2}$ along the whole contour $l$ is nonzero and if the magnetic flux within $l$ is not divisible by the flux quantum $\Phi _{0} = \pi \hbar /e$, i.e. if  $\oint dl A  = \Phi \neq n\Phi _{0}$. The energy density of the superconducting state increases because of non-zero velocity
$$f_{GL} = (\alpha + \frac{mv^{2}}{2})|\Psi |^{2} + \frac{\beta }{2}|\Psi |^{4} \eqno{(3)}$$
and reduces the superconducting transition temperature $T_{c}(v)$ corresponding to $\alpha +mv^{2}/2 = \alpha _{0}[T/T_{c}(v=0) - 1] + mv_{2}/2 = 0$  [5]. The idea of the Little-Parks experiment [1] is based on the dependence of the critical temperature $T_{c}(v) = T_{c}(v=0)[1-mv^{2}/2\alpha _{0}] = T_{c}(v=0)[1-(\xi ^{2}(0)/r^{2})(n-\Phi /\Phi _{0})^{2}]$ of a thin-walled (the wall thickness is smaller than the penetration depth of magnetic field:   $w \ll \lambda _{L}$) superconducting cylinder, whose radius $r$ is comparable with the correlation length $\xi (0) = \hbar ^{2}/2m\alpha _{0}$. The paper by Little and Parks [1] was entitled "Observation of Quantum Periodicity in the Transition Temperature of a Superconducting Cylinder." But in reality the authors observed not the periodicity of the transition temperature $T_{c}(\Phi /\Phi _{0})$ but the periodicity of the resistance $R(\Phi /\Phi _{0})$ of the cylinder at $T \approx  T_{c}$, where $R(T)$ is non-zero, $R(T) > 0$, but smaller than the resistance in the normal state, $R(T) < R_{n}$. This means that the Little-Parks effect is of a fluctuation nature, because, without fluctuations, the superconducting pair density is nonzero, i.e., $|\Psi |^{2} > 0$, only in the superconducting state at $T < T_{c}$ where $R(T) = 0$. Fluctuations in superconductors became the object of intensive research [6] several years after the publication by Little and Parks [1]. Therefore, in [1], and also in [5] (presumably, by tradition), a purely empiric relationship between the periodicities of resistance and critical temperature was employed: $\Delta R(\Phi /\Phi _{0}) \approx  [dR/d(T - T_{c})] \Delta T_{c}(\Phi /\Phi _{0})$ 
\begin{figure}[b]
\includegraphics{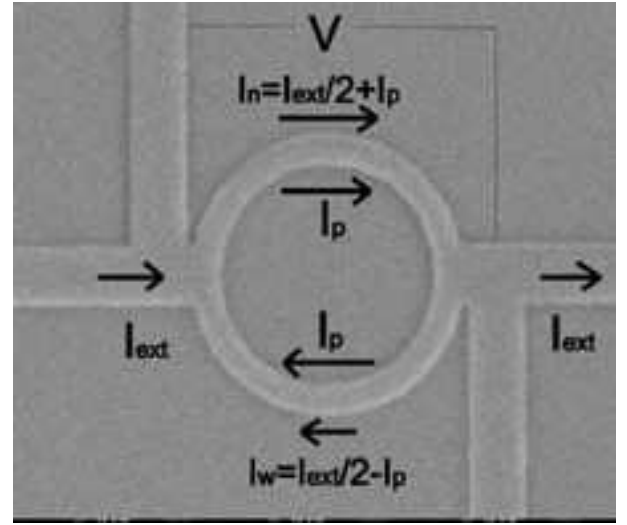}
\caption{\label{fig:epsart} Structure used for the observation of the Little-Parks resistance oscillations $V(\Phi /\Phi _{0}) = V(\Phi /\Phi _{0})/I_{ext}$.}
\end{figure}

The normal state at $T < T_{c}$ and the superconducting state at $T > T_{c}$ have a finite probability $P(\Psi ) \propto exp(-F_{GL}/k_{B}T)$ [5,7] because of the thermal fluctuations. Therefore, the thermodynamic average density of superconducting pairs $\overline{|\Psi |^{2}} \neq 0$ and the circulating persistent current $I_{p} = s2e\overline{|\Psi |^{2}}v \neq 0$ is observed at $\Phi  \neq n\Phi _{0}$ (when the pair velocity $v \neq 0$ according to (2)) in the critical region of the superconducting transition in spite of a non-zero resistance $R_{n} > R(T) > 0$. The displacement $T_{c}(v) = T_{c}(v=0)[1 - mv^{2}/2\alpha _{0}]$ of the resistive transition $R(T)$ observed at $v \neq 0$ results from a change in the probability $P(\Psi )$ of superconducting state $|\Psi |^{2} \neq 0$ (depending on the total energy $F_{GL}$) because of the energy $F_{GL}$ increase at $v \neq 0$.  

\begin{figure}
\includegraphics{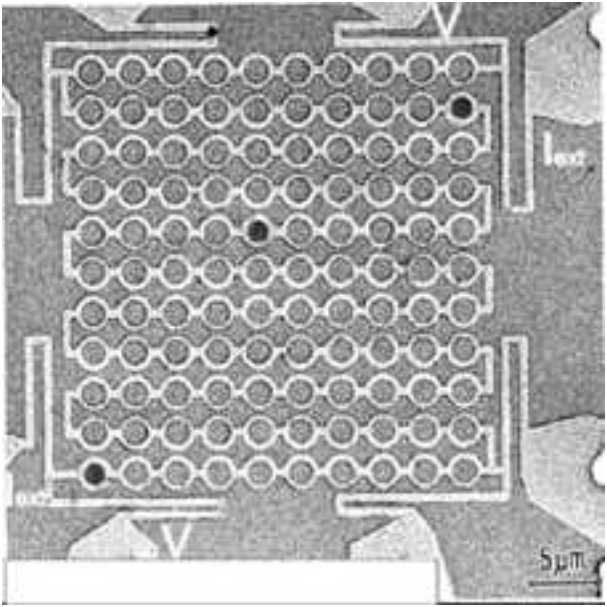}
\caption{\label{fig:epsart} System of 110 series-connected asymmetric alumi-num rings with a diameter of $2 \ \mu m$ and half-ring widths of $0.2 \ \mu m$ and $0.4 \ \mu m$. The voltage was measured between the $V-V$ contact pads, and, for the current, the $I_{ext} - I_{ext}$ contact pads were used. The four additional contact pads, which allowed the measurements on parts of the system, were not used in the experiment under discussion.}
\end{figure}

Thus, the Little-Parks experiment is the first experi-mental evidence of the persistent current $I_{p}$ [8] observed at nonzero resistance $R(T) > 0$ because of the thermal fluctuations. In a symmetric ring with a uniform cross section $s$ and a mean pair density $\overline{|\Psi |^{2}}$  along the circumference $l$, the persistent current is $I_{p} = s(4e^{2}/ml)\overline{|\Psi |^{2}}\overline{(n\Phi _{0}-\Phi )} = I_{p,A}2(\overline{n}\Phi _{0}-\Phi )$ where $I_{p,A} = s(2e^{2}/ml)\overline{|\Psi |^{2}}$ is the amplitude of oscillations of the persistent current in magnetic field for $(\overline{n}\Phi _{0}-\Phi)$ varying from $-1/2$ to $1/2$ [5]. The presence of the persistent current $I_{p}(\Phi /\Phi _{0}) = I_{p,A}2(\overline{n}\Phi _{0}-\Phi )$ at $T > T_{c}$ was theoretically justified in [9, 10] and experimentally confirmed by magnetic susceptibility measurements [11].

It is necessary to note a certain discrepancy between the Little-Parks experiment and its commonly accepted interpretation [5]. The point is that the measured quantity is the voltage $V(\Phi /\Phi _{0})$ at a finite value of the mea-suring current $I_{ext}$ (Fig. 1), whereas the interpretation deals with the resistance oscillations $\Delta V(\Phi /\Phi _{0})/I_{ext} = \Delta R(\Phi /\Phi _{0}) \approx  (-dR/dT)\Delta T_{c}(\Phi /\Phi _{0})$ in the limit $I_{ext} \rightarrow 0$. The discrepancy should be especially significant, if the persistent current $I_{p}$, as the conventional circulating current, has a direction, because, in this case, at $I_{ext} < 2I_{p}$, the total current in one of the semi-rings $I_{w} = I_{ext}/2 - I_{p}$ (or $I_{n} = I_{ext}/2 - I_{p}$) will be directed against the electric field $E = -\nabla V$ (Fig. 1). Because of a possibility of such the paradoxical situation the measurement of the Little-Parks oscillations with the $I_{ext}$ value as small as possible may have fundamental importance.  

\begin{figure}
\includegraphics{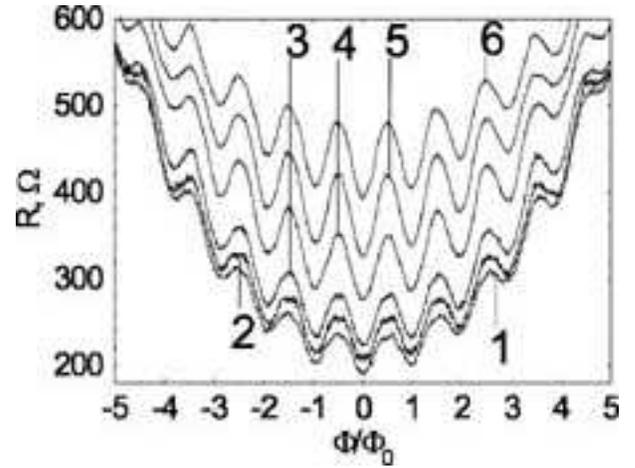}
\caption{\label{fig:epsart} Little-Parks oscillations measured with different values of the current: $I_{ext}$ = (1) 100, (2) 200, (3) 300, (4) 400, (5) 500, and (6) 600 nA, at temperature T = (1) 1.3530, (2, 3) 1.3528, (4) 1.3515,(5) 1.3507, (6) 1.3503 K corresponding to the lower part $R(T) \approx  0.2R_{n}$ of the resistive transition (See Fig. 6).}
\end{figure}

To measure the oscillations $\Delta V(\Phi /\Phi _{0})$ at small $I_{ext}$ values, we used a system consisting of a great number of aluminum rings (Fig.2). In order to prove that the persistent current, like the conventional current, has a direction, we used the system of asymmetric rings. The nanostructure shown in Fig.2 was fabricated by the lift-off method by depositing a thin aluminum film of thickness $d = 20 \ nm$ on a Si substrate. The lithography was performed using a JEOL-840A scanning electron microscope, which was transformed into a laboratory electron lithograph by the NANOMAKER program package. The structure consists of 110 series-connected asymmetric rings with the same inner diameter $2r = 1.9 \ \mu m$ and the semi-ring widths $w_{w} \approx  0.4 \ \mu m$ and $w_{n} \approx  0.2 \ \mu m$ (Fig.2). The cross sections of the semi-rings are $s_{w} = w_{w}d \approx  0.008 \ \mu m^{2}$ and $s_{n} = w_{n}d \approx  0.004 \ \mu m^{2}$. With the London penetration depth in aluminum being $\lambda _{L}(T) \approx  0.05 \ \mu m (1 - T/T_{c})^{-1/2}$, these cross sections correspond to a weak screening $s_{n} < s_{w} < \lambda _{L}^{2}(T)$ at $T > 0.7T_{c}$. The distance between the rings and the width of the strips connecting them is $\approx 0.5 \ \mu m$. The midpoint of the resistive transition corresponds to $T_{c} \approx  1.36 \ K$, the width of the transition is $\Delta T_{c}(0.1 \div 0.9R_{n}) \approx  0.04 \ K$, and the maximum slope is $dR/d(T-T_{c}) \approx  30000 \ \Omega /K$. The resistance of the structure in the normal state is $R_{n} \approx  970 \ \Omega $ the resistance per square is $R_{n} \approx  1.4 \ \Omega /\diamondsuit $, and the resistance ratio is $R(300 \ K)/R(4.2 \ K) \approx  1.7$. 

\begin{figure}
\includegraphics{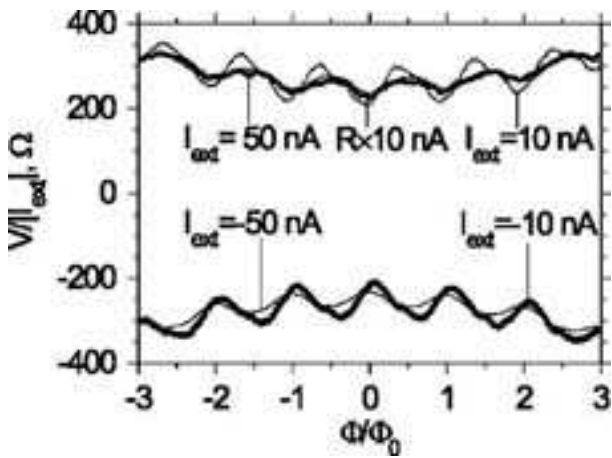}
\caption{\label{fig:epsart} Little-Parks oscillations measured for the opposite directions of the measuring current: $I_{ext}$ = 50, 10, -10, and -50 nA at T = 1.353 K. The increase in the amplitude of the oscillations and the displacement of the extrema at $I_{ext}$ = 10 nA are related to the presence of the voltage $V_{p}(\Phi /\Phi _{0})$ at $I_{ext}$ = 0 (see Fig. 5). The subtraction of this voltage $[V(\Phi /\Phi _{0})-V_{p}(\Phi /\Phi _{0})]/I_{ext}$ leads to the Little-Parks universal dependence at $I_{ext}$ = 10 nA (curve $R \times  10 \ nA$). }
\end{figure}

The measurements were performed by the four-probe method (Fig.2) in a glass cryostat with the use of He4 as the cooling agent. By pumping, the temperature could be reduced to $1.2 \ K$. A constant or sine current generated by a Keithley 6221 precision source was supplied to the current contact pads $I_{ext} - I_{ext}$ (Fig.2). The constant current from $1 \ nA$ to $2 \ \mu A$ was used to measure the resistance versus magnetic field $R(B)$ (the Little-Parks oscillations) and temperature $R(T)$. The sine current was used to obtain the dependencies of the dc voltage on magnetic field $V_{dc}(B)$. The voltage was mea-sured across the potential contact pads $V-V$ (Fig.2) by an instrumental amplifier with a gain of 1000 and an input-normalized noise level of $20 \ nV$ in the frequency band from 0 to 1 Hz. Then, the signal was supplied to an SR560 preamplifier (Stanford Research Systems), which was used for additional amplification and formation of the required signal frequency band by low-pass and high-pass filters. The temperature was measured by a calibrated resistance thermometer ($R(300 K) = 1.5 \ k\Omega $) with a measuring current of $0.1 \ \mu A$. The amplified voltage taken from the sample and the signals that were proportional to the current passing through the sample, to the magnetic field, and to the temperature were simultaneously digitized by a 16-digit A/D converter with eight differential inputs.

\begin{figure}
\includegraphics{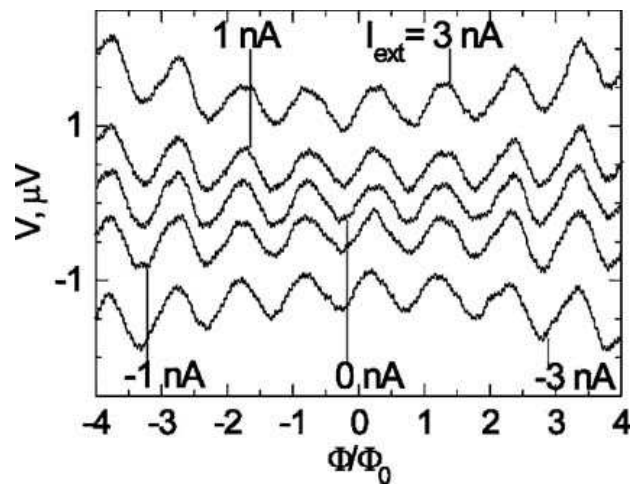}
\caption{\label{fig:epsart} Voltage oscillations $V(\Phi /\Phi _{0})$, measured for different values and directions of the direct current: $I_{ext} = 3, 1, 0, -1,  -3 \ nA$ at $T = 1.358 \ K$. }
\end{figure}

To estimate the persistent current in the critical region $T \approx  T_{c}$, we used the fact that not only the persistent current $I_{p}$, but also the measuring current $I_{ext}$ shifts the resistive transition and causes a resistance variation $\Delta R$. The measurements of $R(T)$ performed by us with measuring currents $I_{ext}$ from $20 \ nA$ to $2 \ mA$ showed that a noticeable displacement $\Delta T_{c}(I_{ext})$ of the resistive transition occurs for $I_{ext} > 100 \ nA$ (at $I_{ext} = 200 \ nA$, $\Delta T_{c}(I_{ext}) \approx  0.001 K$). In accordance with the $T_{c}$ change the resistance increases with increasing $I_{ext}$ when $I_{ext} > 100 \ nA$ (Fig. 3). At $I_{ext} = 300 \ nA$, this increase is close to the amplitude of the $\Delta R(\Phi /\Phi _{0})$ oscillations observed in the experiment (Fig.3). This suggests that the amplitude $I_{p,A}$ of the persistent current oscillations is no smaller than $100 \ nA$ at this temperature, corresponding to the lower part of the resistive transition, see Fig. 6. This $I_{p,A}$ value is approximately an order of magnitude smaller than the value obtained from magnetic susceptibility measurements on a similar ring (at $T \approx  T_{c}$, $I_{p,A}\approx 1 \ \mu A$) [11] and is closer to the theoretical value predicted in [10]. For comparison, according to the results of magnetic susceptibility measurements [11] and the measurements of the critical current oscillations [12,13] in similar aluminum rings the amplitude of the persistent current should be $I_{p,A}\approx 100 \ \mu A (1- T/T_{c})$ in the superconducting state, i.e.  at $T < T_{c}$.

\begin{figure}
\includegraphics{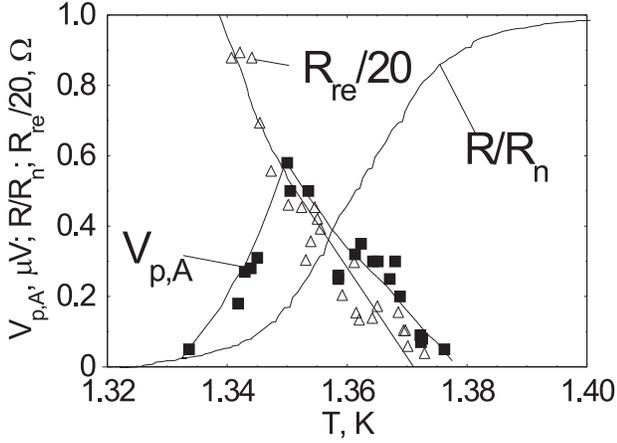}
\caption{\label{fig:epsart} Temperature dependences of the oscillation amplitude $V_{p,A}$ of noise-induced voltage $V_{p}(\Phi /\Phi _{0})$ (the measured values are indicated by squares) and the rectification efficiency $R_{re}$ (indicated by triangles). The data are shown on the background of the resistive transition ($R/R_{n}$).}
\end{figure}

Our measurements showed that the $\Delta R(\Phi /\Phi _{0})$  oscillations are qualitatively independent of the ratio $I_{ext}/I_{p,A}$ of the measuring $I_{ext}$ and the persistent current $I_{p,A} \approx 100 \ nA$ values (Fig. 3). An increase in the $\Delta R(\Phi /\Phi _{0})$ oscillation amplitude from $\Delta R_{A} \approx 50 \ \Omega $ at $I_{ext} = 100 \ nA$ to $\Delta R_{A} \approx 90 \ \Omega $ at $I_{ext} = 600 \ nA$ (Fig. 3) is caused by the difference in the slopes of the resistive transition $dR/d(T- T_{c})$ at $R(T- T_{c}) \approx  0.2R_{n}$ and at $R(T- T_{c}) \approx  0.5R_{n}$ (Fig. 6) and corresponds to approximately equal values of the critical temperature oscillation amplitude $\Delta T_{c,A} = \Delta R_{A}[dR/d(T- T_{c})]^{-1} \approx  0.003 \ K$. The value of $\Delta T_{c,A}/T_{c} \approx  0.0025$ observed in the experiment is close to the $\Delta Tc/Tc = (\xi ^{2}(0)/r^{2})(n-\Phi /\Phi _{0})^{2}$ oscillation amplitude obtained from the theory [5]: $\Delta T_{c,A}/T_{c} = (\xi ^{2}(0)/r^{2})(1/4)  \approx  0.004$ at the ring radius $r \approx  1\ \mu m$  and the correlation length $\xi (0) \approx  130 \ nm$ (which was determined for similar aluminum films in [14]). Despite the fact that we performed the measurements with a system of rings rather than with one ring, the value observed for $\Delta T_{c,A}/T_{c}$ agrees well with the value estimated in the approximation of the lowest permitted state, when the $n-\Phi /\Phi _{0}$ value varies from $-1/2$ to $1/2$.

When $I_{ext} = 50 \div 300 \ nA$, the amplitude $\Delta R_{A}$ varies insignificantly (Fig. 3), and the measured voltage oscillations can be described by the general formula $V(\Phi /\Phi _{0}) = R(\Phi /\Phi _{0})I_{ext} = R_{0}I_{ext} + \Delta R(\Phi /\Phi _{0})I_{ext}$ where $R_{0}$ is the resistance of the ring without the Little-Parks oscillations taken into account. This resistance monotonically increases with magnetic field. The value of $R_{0}(\Phi )$ is approximately the same for $I_{ext} = 50$ and $100 \ nA$; it is approximately 10 and 30 $\Omega $ greater for $I_{ext} = 200$ and $300 \ nA$, respectively (Fig. 3). The minimum of the resistance $R(\Phi /\Phi _{0}) = V(\Phi /\Phi _{0})/I_{ext}$ in the Little-Parks oscillations are observed at $\Phi  = n\Phi _{0}$ and the maximum are observed at $\Phi  = (n+0.5)\Phi _{0}$ (Fig. 3); the ratio $V(\Phi /\Phi _{0})/I_{ext}$ does not depend on the direction of the measuring current. For $I_{ext} < 50 \ nA$, a deviation from the given universal dependence is observed (Fig. 4). The $V(\Phi /\Phi _{0})/I_{ext}$ dependencies measured for opposite directions of $I_{ext}$ are displaced with respect to each other: by $\approx  0.12\Phi _{0}$ at $I_{ext} = 20 \ nA$; by $\approx  0.20\Phi _{0}$ at $I_{ext} = 10 \ nA$ (Fig. 4); by $\approx  0.30\Phi _{0}$ at $I_{ext} = 5 \ nA$; by  $\approx  0.44\Phi _{0}$ at $I_{ext} = 3 \ nA$; and, finally, at $I_{ext} = 1 \ nA$, the displacement $\approx  0.5\Phi _{0}$ reaches the half-period of oscillations (Fig. 5). 

The deviation from the universal dependence at $I_{ext} < 50 \ nA$ is a consequence of an alternating-sign oscillations of the dc voltage $V_{p}(\Phi /\Phi _{0})$ observed in the absence of any external current, i.e., at $I_{ext} = 0$ (Fig. 5). The sign of $V_{p}(\Phi /\Phi _{0})$ changes at $\Phi  = n\Phi _{0}$ and $\Phi  = (n+0.5)\Phi _{0}$; its extrema are observed at $\Phi  = (n \pm 0.25)\Phi _{0}$ (Fig. 5). Because of the presence of alternating-sign $V_{p}(\Phi /\Phi _{0})$ oscillations (Fig. 5), the total voltage is $V(\Phi /\Phi _{0}) = R_{0}I_{ext} + \Delta R(\Phi /\Phi _{0})I_{ext} + V_{p}(\Phi /\Phi _{0})$ where $R_{0}I_{ext} + \Delta R(\Phi /\Phi _{0})I_{ext} = V(\Phi /\Phi _{0}) - V_{p}(\Phi /\Phi _{0})$ corresponds to the Little-Parks universal dependence (Fig. 3) at $I_{ext} = 20$ and $10 \ nA$ (Fig. 4). The voltage amplitude of the Little-Parks oscillation (Fig. 4) is $\Delta R_{A} \times I_{ext} \approx  50 \ \Omega \times 10 \ nA \approx  0.5 \ \mu V$  at $I_{ext} = 10 \ nA$, or $0.5 \ \mu V/110 \approx  5 \ nV$ per one ring. At $I_{ext} = 5$ and $3 \ nA$, the voltage oscillation amplitude is $\Delta R_{A} \times I_{ext} \approx  50 \ \Omega \times I_{ext} \leq   0.25 \ \mu V$. This value only slightly exceeds the noise level and is too small for a reliable observation of the Little-Parks oscillations.

The similarity of the  $V_{p}(\Phi /\Phi _{0})$ oscillations (Fig. 5) to the oscillations of the thermodynamic average $\overline{I_{p}} \propto  \overline{v} \propto  (\overline{n}-\Phi /\Phi _{0})$ indicates that the persistent current $I_{p}$, like the conventional circulating current $V = (R_{srn} - R_{srw})I = R_{as}I$, causes a potential difference $V_{p}(\Phi /\Phi _{0})  = R_{ef}\overline{I_{p}} (\Phi /\Phi _{0})$ on the semi-rings with different cross sections $w_{w} \approx  0.4 \ \mu m \neq w_{n} \approx  0.2 \ \mu m$ (Fig. 2). Periodic changes of the direction of electric field $E = -\nabla V_{p}(\Phi /\Phi _{0})$ with varying magnetic flux value $\Phi /\Phi _{0}$  (Fig. 5) prove that the persistent current has clockwise or contra-clockwise direction, like conventional current. 

Presumably, the voltage $V_{p}(\Phi /\Phi _{0})$ is induced by noise, which is present in any real measuring system. The amplitude $V_{p,A}$ of the $V_{p}(\Phi /\Phi _{0})$ oscillations depends nonmonotonically on temperature and reaches its maximum $maxV_{p,A} \approx  0.6 \ \mu V$ at $T \approx  1.350 \ K$, which corresponds to the lower part of the resistive transition $R(T) \approx  0.15R_{n}$ (Fig. 6). To estimate the intensity of noise inducing $V_{p}$, we measured the amplitude $V_{A}$ of the oscillations of the dc voltage $V_{dc}(\Phi /\Phi _{0})$ as a function of the amplitude $I_{0}$ of the external ac current $I_{ac} = I_{0}\sin (2\pi ft)$ at different temperatures. In [15], it was shown that the $V_{A}$ value does not depend on the frequency $f$ of the current in a wide frequency range and that it nonmonotonically depends on the amplitude $I_{0}$ by reaching its maximum value $V_{A,max}$ at $I_{0} = I_{0,max}$ close to the critical current at a given temperature. The efficiency of rectification $R_{re} = V_{A,max}/I_{0,max}$ increases with an increase in the number of rings in the system; at a sufficiently low temperature it is approximately equal to one fifth of the resistance in the normal state $R_{re} \approx   0.2R_{n}$ independently of the ring number in the system and its resistance $R_{n}$ [13]. For our system of 110 rings with $R_{n} \approx  970 \ \Omega $, we observed the $V_{dc}(\Phi /\Phi _{0})$ oscillations with an amplitude reaching $V_{A,max} \approx  0.0045 \ V$, and $R_{re} \approx  180 \ \Omega \approx  0.2R_{n}$ at the low temperature $T \approx  1.20 \ K$. Near the beginning of the transition to the normal state, the rectification efficiency decreases to $R_{re} \approx  20 \ \Omega \approx  0.02R_{n}$, and the decrease continues as the temperature increases further (Fig. 6). Above $T \approx  1.350 \ K$ corresponding to $maxV_{p,A} \approx  0.6 \ \mu V$ the quantity $R_{re}$ decreases simultaneously with $V_{p,A}$ (Fig. 6), and, on the average, $V_{p,A}/R_{re} \approx  50 \ nA$. This value estimates the noise amplitude in our system: $I_{0,noise} \approx  50 \ nA$. Note that the effective noise power equal to $(R_{n}/110) \overline{I_{noise}^{2}} \approx 2 \ 10^{-14} \ W$ per one ring, corresponds to the wide frequency spectrum within which the rectification effect is observed. For comparison, the total equilibriom power of the Nyquist noise $W_{Nyq} = k_{B}T\Delta \omega  = (k_{B}T)^{2}/\hbar  \approx  3 \ 10^{-12} \ W$ at $T = 1.3 \ K$ in the whole frequency band from 0 to the quantum limit $k_{B}T/\hbar \approx 1.8 \ 10^{11} \ Hz$ is two orders of magnitude greater. This estimate, as well as the decrease in $V_{p,A}$ to zero in the lower part of the resistive transition (Fig. 6), testifies to a low noise level in our system and suggests that a system of asymmetric superconducting rings can be used as a detector of this kind of noise [16]. 

Thus, the use of the system with a great number of rings allowed us to observe the Little-Parks oscillations with a measuring current $I_{ext}$ much smaller than the amplitude of the persistent current $I_{p,A}$. This result sug-gests that the Little-Parks oscillations $R(\Phi /\Phi _{0})$ should be observed in the limit $I_{ext} \rightarrow  0$, i.e. under equilibrium, which confirms the assumption used in the explanation of this effect [5]. The use of asymmetric rings allowed us to show that the persistent current has a direction and that this direction periodically changes as the magnetic field varies. This means that, when $I_{ext}/2 < I_{p,A}$, the constant component of the total current in one of the semi-rings is directed against the dc electric field $E = -\nabla V$, because the persistent current, unlike the conventional circulating current, is observed in the absence of the Faraday electromotive force, i.e. at $d\Phi /dt = 0$. To reveal the cause and nature of this paradox, additional investigations are necessary.

\section*{Acknowledgement}
This work was supported by the Presidium of the Russian Academy of Sciences (program "Quantum Nanostructures") and by the Division of Information Technologies and Computational Systems of the Russian Academy of Sciences (project "A Quantum Bit on the Basis of Micro- and Nanostructures with Metallic Conduction" in the program "Organization of Calculations by Using New Physical Principles").


\begin{thebibliography}{99}

\bibitem{LP1962} W. A. Little and R. D. Parks, {\em Phys. Rev. Lett}. {\bf 9}, 9 (1962).

\bibitem{FlQu1961} B. S. Deaver and W. M. Fairbank, {\em Phys. Rev. Lett}. {\bf 7}, 43 (1961); R. Doll and M. Nabauer, {\em Phys. Rev. Lett}. {\bf 7}, 51 (1961).

\bibitem{GL1950} V. L. Ginzburg and L. D. Landau, {\em Zh. Eksp. Teor. Fiz}. {\bf 20}, 1064 (1950).

\bibitem{London50} F. London, {\em Superfluids} (Wiley, New York, 1950), Vol. 1.

\bibitem{Tinkham}  M. Tinkham, {\em Introduction to Superconductivity} (McGraw- Hill, New York, 1975; Atomizdat, Moscow, 1980).

\bibitem{AL1968} L. G. Aslamazov and A. I. Larkin, {\em Fiz. Tverd. Tela} (Leningrad) {\bf 10}, 1104 (1968) [{\em Sov. Phys. Solid State} {\bf 10}, 875 (1968)]; {\em Phys. Lett. A} {\bf 26}, 238 (1968).

\bibitem{SkTi1975} W. J. Skocpol and M. Tinkham, {\em Rep. Prog. Phys}. {\bf 38}, 1049 (1975).

\bibitem{Blatt61} J. M. Blatt, {\em Phys. Rev. Lett}. {\bf 7}, 82 (1961).

\bibitem{Kulik70}  I. O. Kulik, {\em Zh. Eksp. Teor. Fiz}. {\bf 58}, 2171 (1970) [{\em Sov. Phys. JETP} {\bf 31}, 1172(1970)].

\bibitem{PerCurTe}  F. von Oppen and E. K. Riedel, {\em Phys. Rev. B} {\bf 46}, 3203 (1992)

\bibitem{PerCurEx} X. Zhang and J. Price, {\em Phys. Rev. B} {\bf 55}, 3128 (1997) 

\bibitem{JETP07Ju} V. L. Gurtovoi, S. V. Dubonos, S. V. Karpii, et al., {\em Zh. Eksp. Teor. Fiz}. {\bf 132}, 297 (2007) [{\em JETP} {\bf 105}, 262 (2007)].

\bibitem{JETP07De} V. L. Gurtovoi, S. V. Dubonos, A. V. Nikulov, et al., {\em Zh. Eksp. Teor. Fiz}. {\em 132}, 1320 (2007) [{\em JETP} {\bf 105}, 1157 (2007)].

\bibitem{Mosh1992}  H. Vloeberghs, V. V. Moshchalkov, C. Van Haesendonck, et al., {\em Phys. Rev. Lett.} {\bf 69}, 1268 (1992).

\bibitem{Lett2003} S. V. Dubonos, V. I. Kuznetsov, I. Zhilyaev, et al., {\em Pis'ma Zh. Eksp. Teor. Fiz.} {\bf 77}, 439 (2003) [{\em JETP Lett.} {\bf 77}, 371 (2003)].

\bibitem{SPIE2006} V. L. Gurtovoi, S. V. Dubonos, A. V. Nikulov, et al., {\em Proc.SPIE} {\bf 6260}, 62600T1 (2006).

\end{thebibliography}
\end{document}